Magnetic levitation within a microwave cavity: characterization, challenges, and possibilities

Nabin K. Raut[1, a)], Jeffery Miller[1], Raymond E. Chiao[1], Jay E. Sharping[1]

University of California, Merced

5200 N. Lake Road, Merced, CA 95343

a) Author to whom correspondence should be addressed: nraut@ucmerced.edu

The low energy losses in the superconducting magnetic levitation make it attractive for exciting applications in physics. Recently, superconducting magnetic levitation has been realized as novel mechanical transduction for the individual spin qubit in the nitrogen-vacancy center [1]. Furthermore, the Meissner has been proposed for the study of modified gravitational wave detection [2]. Meissner levitation within the microwave cavity could open avenues for the novel cavity optomechanical system, readout for quantum object such as the transmon, and magnon, gravitational wave detection, and magnetomechanics [3]. This work characterized magnetic levitation within a microwave. It also discusses possibilities, challenges, and room temperature and cryogenic experiments of the cavity-magnet system.

## 1. Introduction

A magnet placed above a superconductor induces supercurrent on the surface of the superconductor, screening the magnetic flux from its core [4] (see Fig. 1). This supercurrent produces an opposing magnetic field inside the superconductor which gives rise to a force called a diamagnetic force [5]. When the strength of the diamagnetic force is large enough to balance the force due to gravity, the magnet will levitate above the superconductor. The phenomenon is called magnetic levitation [6].

The magnetic levitated system has been used for the development of the transportation system. For example, the levitated system is implemented in the high-speed train known as the bullet train and in the much-awaited transportation development called the hyperloop. The energy loss is low in such a freely floating system. There is no loss due to clamping and thermal contact. Moreover, the superconducting magnetic levitation further minimizes the losses. There are no losses, as in the optical levitation, due to photon recoil and heating in the superconducting magnetic levitation [7].

A coaxial quarter wave microwave cavities can localize electric or magnetic fields in a small area (see Figure 2) [8--10]. The electric field is localized on the stub and the magnetic field on the bottom of the cavity. Such concentration of the energy in a small volume increases the mode volume of the cavity. This causes a reduction in energy dissipated due to the conductive and dielectric losses at surfaces and interfaces [11]. The quality factor (Q) of the cavity is the ratio of total energy it can store to its total losses. To have a high value of Q indicates a lower energy dissipation, meaning a photon recycles multiple times inside the cavity before dying out. Such a narrow mode of the cavity provides unique opportunities to couple other objects and systems [12,13]. For instance, cavities with higher values of Q are applied to detect rare gravitational wave events [14], dark matter detection [15--19], mechanical coupling, and perturbation [20].

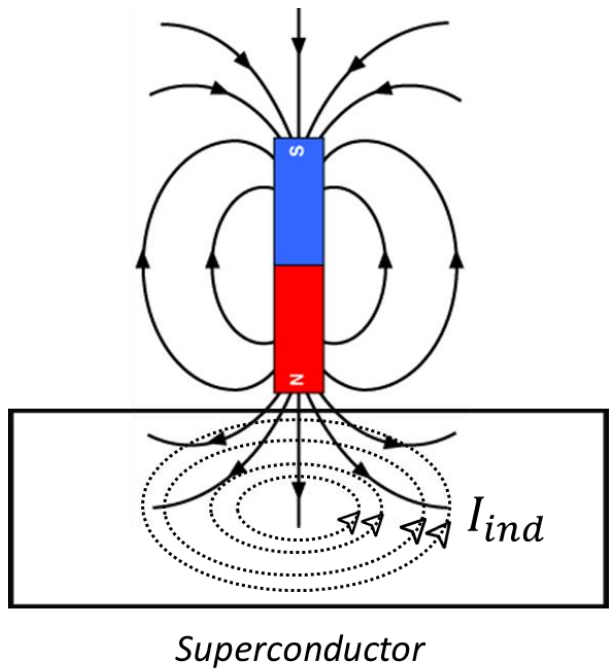

FIG. 1: (a) A magnet levitating above a superconductor because of induced opposing current on the surface of the superconductor, (b) snapshot of a cavity with a magnet attached on the base plate of a dilution refrigerator ready for the cooldown.

The superconducting microwave cavities are fabricated from materials such as aluminum, niobium, etc. Cavities made up of such materials go into the superconducting phase transition at the transition temperature. Large numbers of Cooper pairs are formed during this phase change. The Cooper pairs are special pairs of electrons whose interaction is mediated by the phonon. They exhibit attractive interaction, thereby reducing the total energy of the system. The reduction of energy brings more stability to the system. The superconducting microwave cavities have about five orders of magnitude lower losses than normal conducting cavities [21].

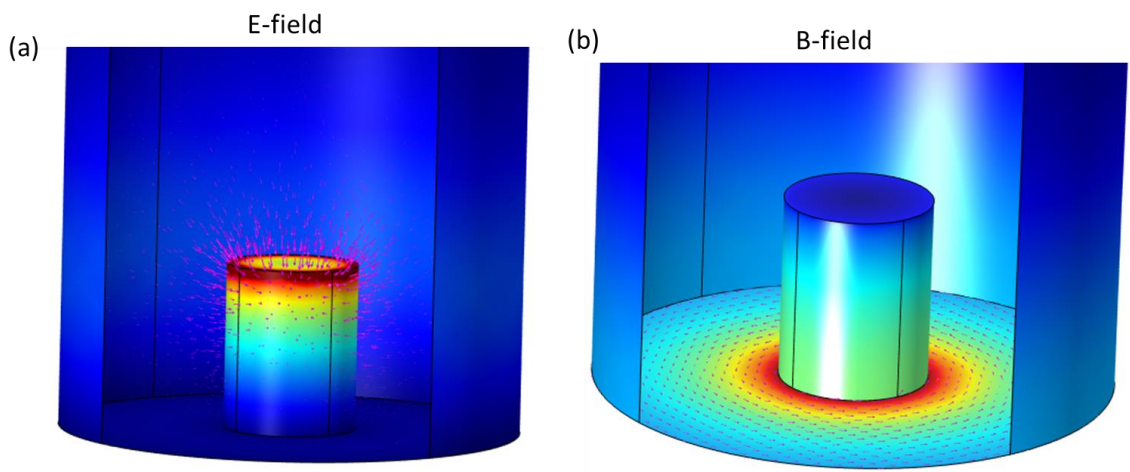

FIG. 2: A coaxial quarter-wave microwave cavity. (a) the electric field is localized around the rim of the stub, (b) the magnetic field is stored on the bottom of the cavity.

One important application of such a high-Q cavity would be to couple it with another oscillator. The coupling here refers to the energy transfer between the two systems. This field of study is commonly known as cavity optomechanics. It studies the dynamics between two coupled harmonic oscillators: a cavity mode and a normal mechanical mode [22]. The field was studied early on theoretically by Braginsky in 1967 [23]. As shown in Figure 3, the circulating field inside the cavity with frequency $\omega$ and damping K generates high power. The enhancement in a photon's momentum can modify the dynamics of the mechanical element [24]. Depending upon the strength of the coupling, rich physics can be explored from semi-classical to novel physics.

Edward Mills Purcell, in 1940s, studied the effect of coupling between the two harmonic oscillators. He observed that the spontaneous emission rate of the quantum system can be enhanced linearly by the quality factor of the cavity. The effect is known as the Purcell effect [25]. Motional cooldown of the classical system to its ground state energy can be achieved with the strong coupling between the two systems. In the quantum mechanics, the least energy state with energy $\frac{1}{2}\hbar\omega$ is called the ground state energy. The ground state cooldown leads to the study of the macroscopic quantum mechanics. In the deep-strong coupling regime the exchange of energy between the light and the matter is faster compared to their losses. Faster exchange of energy caused splitting in the energy level [20].

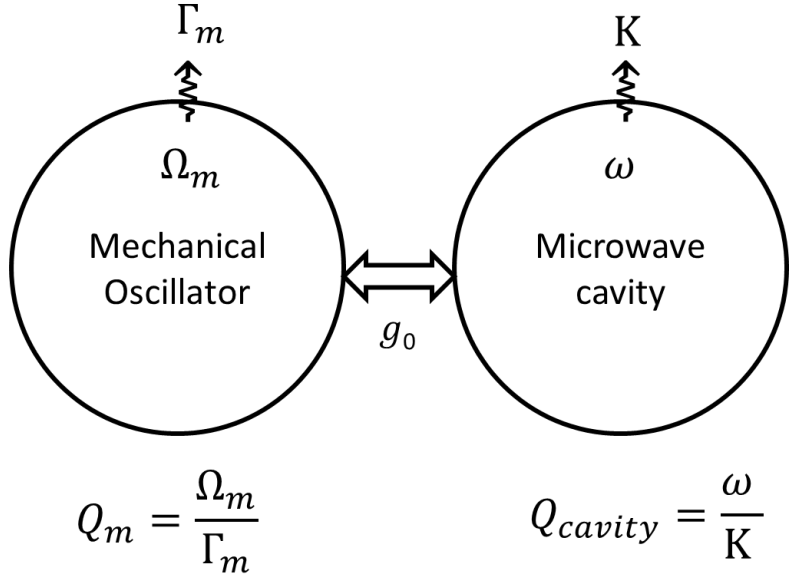

FIG. 3: Schematic of a cavity optomechanical system. Here, a mechanical degree of freedom is coupled to the resonance mode of a microwave cavity. The strength of coupling between two systems is represented by $g_0$. The frequency and damping of the microwave cavity are $\omega$ and K, respectively. Similarly, the mechanical system has the frequency of $\Omega_m$ and the damping of $\Gamma_m$. The quality factor of the mechanical oscillator and the microwave cavity are denoted, respectively, by $Q_m$ and $Q_{cavity}$.

The superconducting microwave cavity is also used for quantum memory. The cavity mode can be used for the control, store, and readout of the quantum states [26]. Figure 4 shows an example of a microwave cavity mode coupling with a qubit mode. When there is no coupling between the modes, one mode's dynamics do not affect the other mode's dynamics. However, when the coupling between the modes is strong, the degeneracy in the qubit energy level lifts. The amount of energy splitting depends directly on the strength of the coupling. Such splitting allows the excitation of the qubit to live for a longer time. Here, the dissipation of the qubit is recycled through cavity mode. Recently, a novel microwave photon counting technique for the detection of low mass bosonic dark matter candidates was developed by using a dispersively coupled superconducting cavity-qubit system [27].

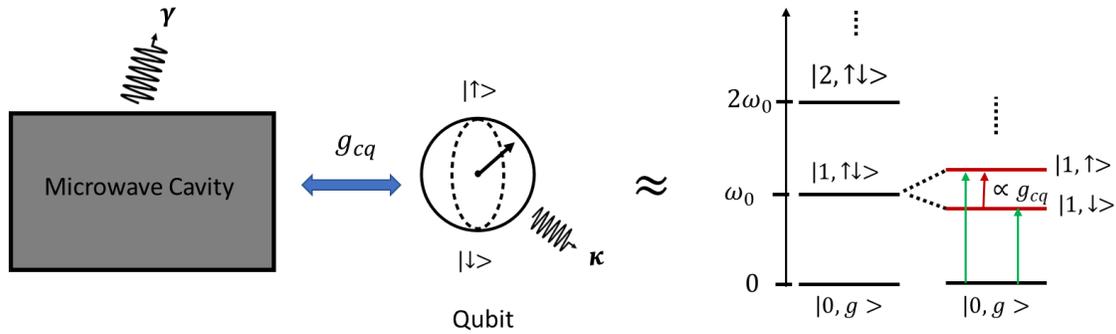

FIG. 4: A high-Q microwave cavity is strongly coupled to a qubit. The strong coupling between the microwave mode and qubit mode results in energy splitting. Here, $\gamma$ and $\kappa$ are the microwave and qubit dissipations, respectively. $g_{cq}$ represents the coupling between the two modes.

Furthermore, interesting results by coupling a mechanical system with the highly localized field of the cavity have been reported. An experimental demonstration of the Casimir spring effect within a superconducting reentrant microwave cavity system was published [28]. Moreover, a strong coupling between a magnon with the highly concentrated RF magnetic field of a cavity has been demonstrated [29,30].

As discussed above, there are promising developments in coupling mechanical oscillators to the superconducting microwave cavity mode [31]. The common feature of all of those perturbations and couplings is they are somehow clamped and are in thermal contact with the other object. This introduces additional loss in the system. Levitated systems are free from those losses. There are significant developments in optical levitation. One main challenge in this type of levitation is the loss associated with the photon recoil and heating [32--34].

A promising alternative to optical levitation is to use passive levitation techniques involving magnets and superconductors. In passive levitation, a magnet is levitated due to the Meissner effect. In this levitation, there are no losses due to clamping, thermal contact, and photon recoil. However, in this technique, it is important to distinguish between Type-I and Type-II superconductors. In the meantime, it is also important to distinguish whether the experiment is performed in a zero-field cooled, or non-zero-field cooled condition. When a Type-II superconductor is cooled below its critical temperature in the presence of a non-zero magnetic field, magnetic flux is trapped by vortices within the material. The trapped flux then freezes the motion of the magnet. In the non-zero-field cooling condition, the Type-II superconductor is cooled in the absence of magnetic fields. The permanent magnet is then inserted after cooling.

In recent research, magnetic levitation above a type-II superconductor has been realized as novel mechanical transduction for the individual spin qubit in the nitrogen-vacancy center [35]. Furthermore, the Meissner has been proposed for the study of modified gravitational wave detection [2].

In comparison, when a Type-I material makes the superconducting transition, the superconductor exhibits a perfectly diamagnetic response as long as the magnetic field is less than some critical value. Supercurrents build up on the surface of the superconductor which fully screens the magnetic flux from its interior [36]. The Meissner force is always repulsive and can be large enough to lift macroscopic objects to some equilibrium height where the net force of the interaction combined with that due to gravity is zero.

Magnetic levitation within a superconducting microwave cavity could be a versatile research platform for tests of fundamental and new physics. First, such an electromechanical system may be useful as a means to couple the low-frequency mechanical motion of the magnet with other quantum objects, such as magnons and transmons, which are used for quantum information processing. The development of sensors based on levitated magnets within high-Q cavities is an interesting alternative for detecting gravitational waves and dark matter. Such a system can be used to prepare, control, store, and measure arbitrary quantum states. Finally, it may be possible to achieve strong coupling to and ground-state cooling of a mechanical resonator.

In this paper, characterization of magnetic levitation within a superconducting microwave cavity has been done. FEM simulation is used to identify possible phenomena that could occur during levitation. Moreover, the quality factor has been discussed as an alternative tool for that purpose. Stable levitation is one of the main challenges in the levitation experiment. A hybrid model is developed to calculate the potential energy of the cavity-magnet system. Possible configurations of the cavity, as well as a magnet, have been discussed in brief to achieve stable levitation. Finally, room temperature and cryogenic experiments are presented and discussed.

## 2. Coaxial Quarter-Wave Cavity

In the cavity configuration being discussed in this work (Fig. 5 (a)), an open-ended cavity is constructed with a stub whose height can be approximated as $\frac{\lambda}{4}$. The electromagnetic mode of interest is the standing wave confined to the bottom, coaxial, section of the cavity (Fig. 5 (b)). The $E$ and $B$ fields oscillate with frequency $\omega$ and resemble a transverse electromagnetic mode. The E field extends radially from the central stub towards the outer wall, while the $B$ field is directed azimuthally around the stub. Another feature of this mode is that the $E$ and $B$ field maxima are localized in different portions of the cavity. The $E$ field amplitude is largest at the top of the coaxial section while the $B$ field is strongest at the base of the stub because of the boundary conditions which state that at the surface of the metal, $H_\perp = 0$ and $E_\parallel = 0$.

In the fundamental mode of the cavity electric field is highly localized around the tip of the stub (Fig. 5 (b)). This non-Maxwellian mode can be approximated as $E_r \propto (E_0 e^{\beta r} + E_0{'} e^{-\beta z}, \; r <$

$r_{stub}$, & $\frac{E_0}{r}\cos(k_0 r)$, $r \geq r_{stub}$), here $r$ is the radial distance and $r_{stub}$ is the radius of the stub [37,38]. Similarly, the axial electric field exponentially decays from the stub towards the open end of the cavity as $E_z \propto E_0 e^{i(\omega t - kz)}$ [39]. Fig. 5 (c) and (d) show COMSOL simulations of the radial and the vertical distribution of the electrical field. Most of the electric field of the cavity is concentrated in the region $1.5\ mm < x < 2.5\ mm\ \&\ 0\ mm \leq z < 1\ mm$. The electric field can further be confined around the stub by changing its shape to the conical. Due to this unique feature, any external perturbation in this region of the cavity results in a significant shift in the resonance frequency.

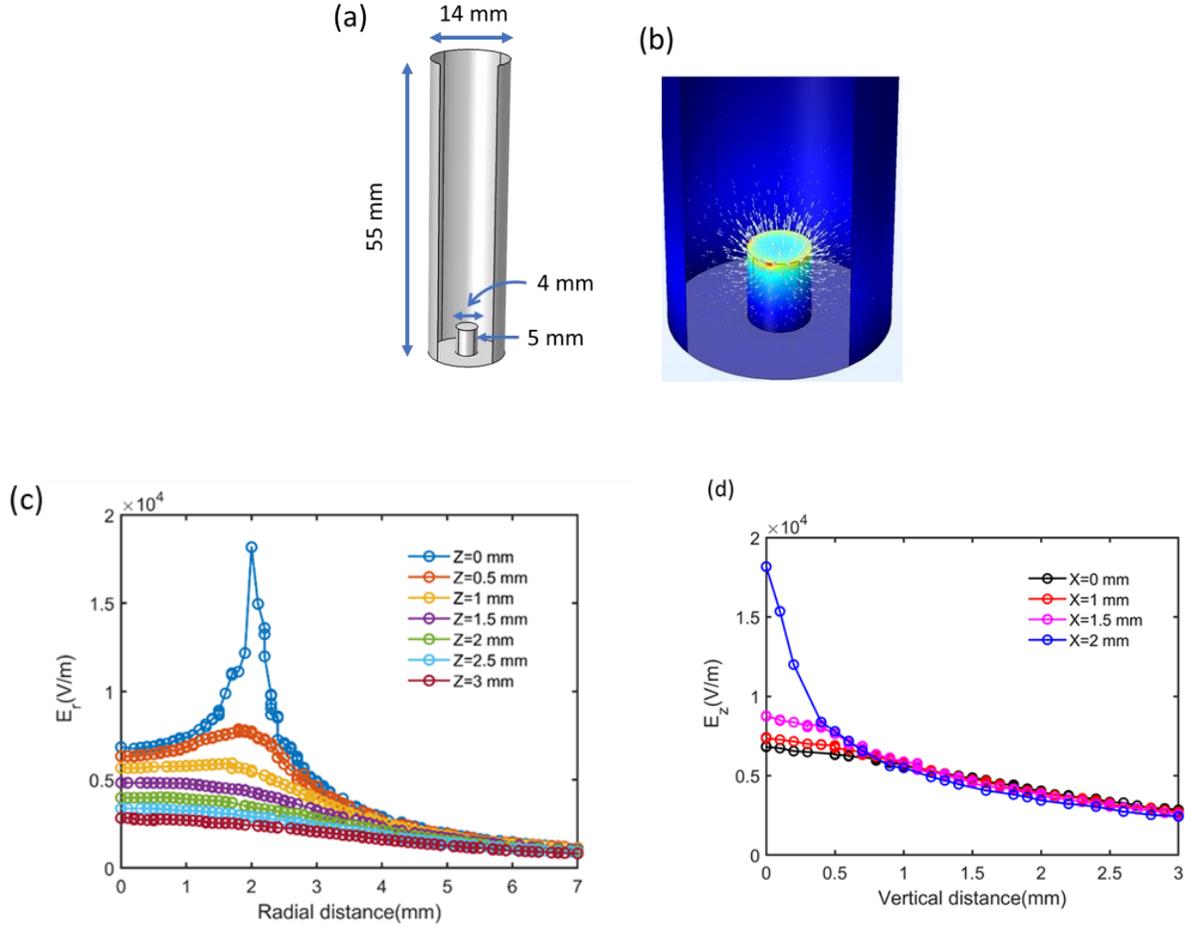

FIG. 5: (a) Cavity configuration, (b) the electric field distribution within the cavity, (c) radial electric field is calculated at a different height from the stub, (d) axial electric field at different radial positions on the stub.

### 3. Neodymium Magnets

The neodymium permanent magnets used in the simulation, analytical calculations, and experiments in this paper are graded according to their remanence ($B_r$) as tabulated in Table 1. For example, the magnet with remanence 1.32 T is called an N42 magnet. The density of the neodymium is $7.4 \times 10^{-3}\ kg/m^3$. The thermal expansion (0 to 100°C) of the neodymium magnet parallel and perpendicular to the direction of magnetization is, respectively, $5.2 \times 10^{-6}$ $1/°C$ and $-0.8 \times 10^{-6}$ $1/°C$ [40].

Table 1: Neodymium magnets classified with their remanence.

| Type of Neodymium magnet | Remanence (T) |
|---|---|
| N35 | 1.22 |
| N38 | 1.26 |
| N40 | 1.29 |
| N42 | 1.32 |
| N45 | 1.37 |
| N48 | 1.42 |
| N50 | 1.44 |
| N52 | 1.47 |
| N54 | 1.50 |

Figure 6 (a) shows a sketch of a disc magnet that has a radius of $R$ and thickness of $2b$. The mass of the magnet ($M$) can be calculated using the relation:
$$M = 2\pi R^2 b \rho \tag{1}$$
Where $\rho$ is the density of the magnet. Figure 6 (b) shows an example of magnetic field lines of an N50 permanent magnet with $R = 0.5 \; mm$ and $b = 0.25 \; mm$. Here (0,0) is the center of the magnet.

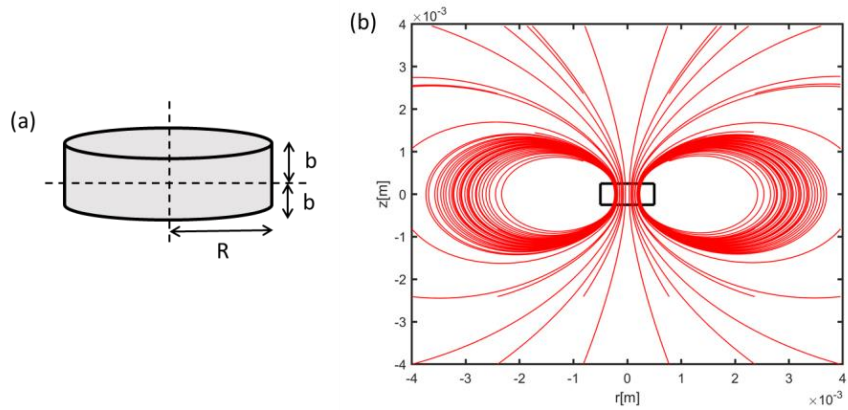

FIG. 6: (a) A disc magnet of a radius of $R$ and a height of $2b$, (b) magnetic field distribution of an N50 magnet.

A simple approach to calculating a permanent magnet's magnetic field is to consider it as a solenoid (loop of currents). Then the magnetic field through the symmetric axis of the magnet is given by [41]:
$$B_z = \frac{B_r}{2} \left[ \frac{Z+b}{\sqrt{R^2 + (Z+b)^2}} - \frac{Z-b}{\sqrt{R^2 + (Z-b)^2}} \right] \tag{2}$$
Here, $B_r$ is a remanence of the magnet. Equation (2) calculates the magnetic field more accurately outside of the magnet than its inside.

Figure 7 shows the decay of the magnetic field for an N50 neodymium magnet of a radius of 0.5 *mm* and height of 0.5 *mm* from its surface. The field varies rapidly from the surface to the distance of 1 mm. For the magnetic field calculation, the origin of the coordinate is considered at the center of the magnet. The strength of the magnet is reduced by approximately half at the surface (~0.5 T) than that at the center of the magnet (1.44 T) and is negligible above 2.5 *mm*.

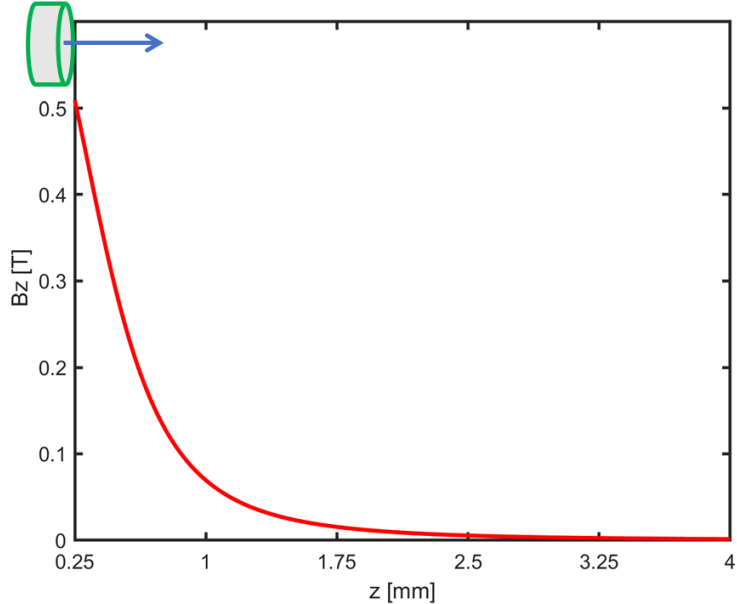

FIG. 7: Azimuthal magnetic field ($B_z$) along the symmetric axis of a magnet of radius 0.5 mm and height 0.5 mm. The field is calculated from the surface of the magnet.

### 4. Magnetic Levitation: Characterization

When a magnet is dropped within the cavity, it affects two crucial aspects of the cavity. First, it interacts with the mode of the cavity. This interaction results in a change in the resonance frequency of the cavity. The second effect is more microscopic. The onset magnetic field can provide external energy to some fraction of the cooper pair near the surface of the superconductor. Thereby delaying superconducting phase transition during cooldown or even destroying the cooper pairs. This can be seen in the quality factor of the cavity. This section of the paper discusses those two effects in brief.

A. FEM Simulations

FEM simulations help understand how a magnet sitting at different locations within the cavity perturbs the mode of the cavity. This effect can be studied by looking into a shift in the resonance frequency of the cavity.

Here, the frequency shift pattern of a cavity magnet is studied by putting a magnet on the different locations around the stub. The magnet used in the simulations is a disc magnet with a diameter of 1 mm and a height of 0.5 mm. Here, simulations for various possible phenomena during the levitation experiment such as magnetic levitation, levitation with an angle, sliding, and flipping of the magnet are studied.

I. Levitation

Figure 8 shows the colormap of the frequency shift as a function of the radial and vertical position of the magnet. The vertical distance is measured from the magnet's surface to the stub's surface and the radial distance from the center of mass of the magnet. For example, when the magnet is at $(x, z) = (2\ mm, 0.5\ mm)$ this means that half part of the magnet is on the stub, while the other half is outside the stub at the height of 0.5 mm. Moreover, the frequency of the bare cavity is 11.007 GHz. The change in the frequency of the cavity ($\Delta f$) is calculated in comparison to the frequency of the bare cavity. The frequency sensitivity refers to the difference in the frequency with respect to the position of the magnet.

In Fig. 8, we can divide the cavity into three regions based on the frequency sensitivity: (a) Near the center of the stub ($x < 1\ mm$), (b) Around the stub ($1\ mm \leq x \leq 3\ mm$), and (c) In the gap of the cavity ($x > 3\ mm$). In region (a) the change in the frequency of the cavity is small with the radial and vertical position of the magnet and more so in the region (c). The region (b) is the most sensitive region of the cavity. In this region, the small displacement of the magnet results in a significant change in the frequency. For example, when the magnet at position ($x = 1.9\ mm, z = 0\ mm$) is moved to the positions ($x = 2\ mm, z = 0\ mm$) and ($x = 2.1\ mm, z = 0\ mm$) the frequency sensitivity will be around 200 MHz/mm. The frequency sensitivity is even high due to the vertical displacement. For example, when the magnet at positions ($x = 1.9\ mm, z = 0\ mm$), ($x = 2\ mm, z = 0\ mm$), and ($x = 2.1\ mm, z = 0\ mm$) is moved to the positions ($x = 1.9\ mm, z = 0.1\ mm$), ($x = 2\ mm, z = 0.1\ mm$), and ($x = 2.1\ mm, z = 0.1\ mm$) the frequency of the cavity is shifted by the rate of 700 MHz/mm.

The change in the frequency ($\Delta f$) of the cavity mainly depends upon the interaction of the magnet with the electric field ($E(r, z)$). In addition, with the amount of electric field storage in the space between the magnet and the stub. Remember, the electric field of the coaxial microwave cavity is localized around the circumference of the stub. Hence, region (c) is the most sensitive region in the entire cavity.

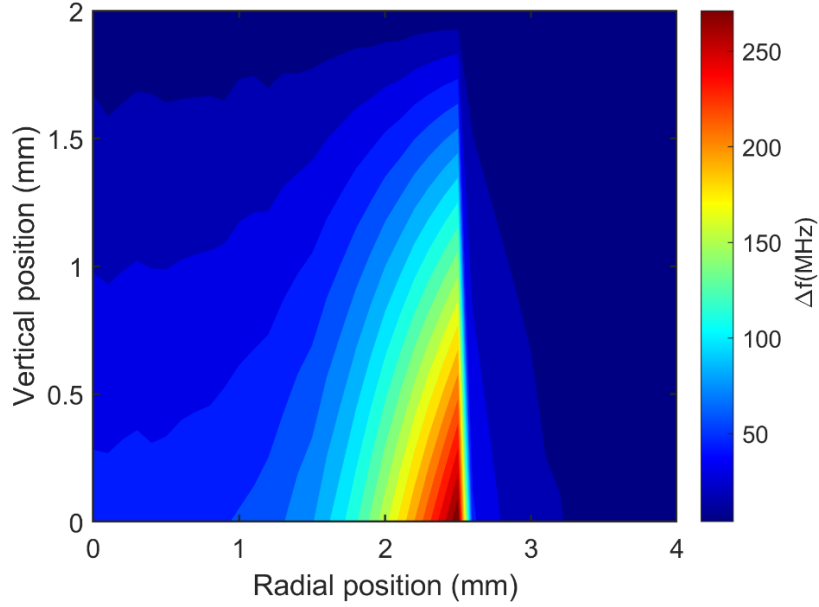

FIG. 8: Colormap of the change in the resonance frequency of the cavity as a function of the radial and vertical position of a magnet on the stub.

II. Levitation with an angle

Suppose a magnet magnetized along its thickness is placed above a superconductor. It will make an angle with the superconductor during levitation to be in the minimum potential energy state [42]. Fig. 9 illustrates the tendency of a magnet to change the resonance frequency of the cavity lifting with an angle from the edge of the stub ($x = 2\ mm$). See the inset of Fig. 9 for the example demonstrations of magnetic levitation for angles $0°, 45°, 90°$. On the surface of the stub, even $10°$ angle will induce the frequency downshift of 30 MHz. The decreasing trend of the frequency will continue as the angle increase from $0°$ to $90°$. However, the trend will reverse, and the frequency of the cavity will start to upshift as the angle change from $90°$ to $180°$.

Magnetic levitation always upshifts the frequency of the cavity, irrespective of the position and orientation of the magnet on the stub. The orientation and position of the magnet play the main role in the amount of the upshift. In Fig. 9, the frequency sensitivity is largest near the stub ($z \sim 0.5\ mm$).

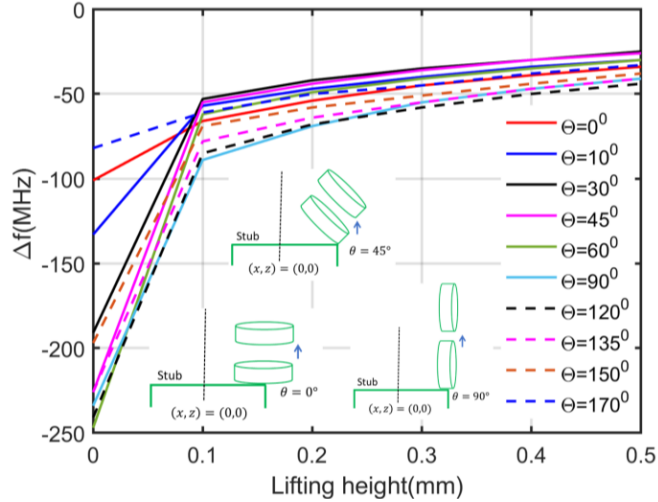

FIG. 9: Magnetic levitation shifts the cavity frequency towards the bare cavity frequency. This plot identifies the amount and trend of the frequency shift for magnetic levitation for different magnet orientations.

III. Sliding towards the edge

Now imagine the following scenario: a magnet that is put at the center of the stub slide towards the edge. One question that might arise is what effect this phenomenon has on the cavity's frequency. In fact, a more general question could be the frequency shift patterns due to the sliding of the magnet.

A perfect diamagnet pushes a magnet towards its edge due to the edge effect [43]. Fig. 10 shows how the frequency shift as a function of radial position for magnets of several angular orientations, as obtained from the FEM simulations. The magnet remains in contact with the surface of the stub for these calculations. The cavity's resonance frequency decreases when a magnet in contact with the stub slides towards the edge. The amount of such a downshift depends on the orientation of the magnet. The sudden changes in the frequency shift as the radial position increases beyond 1.5 mm are due to the edge of the magnet closest to the stub surface extending past the edge of the stub.

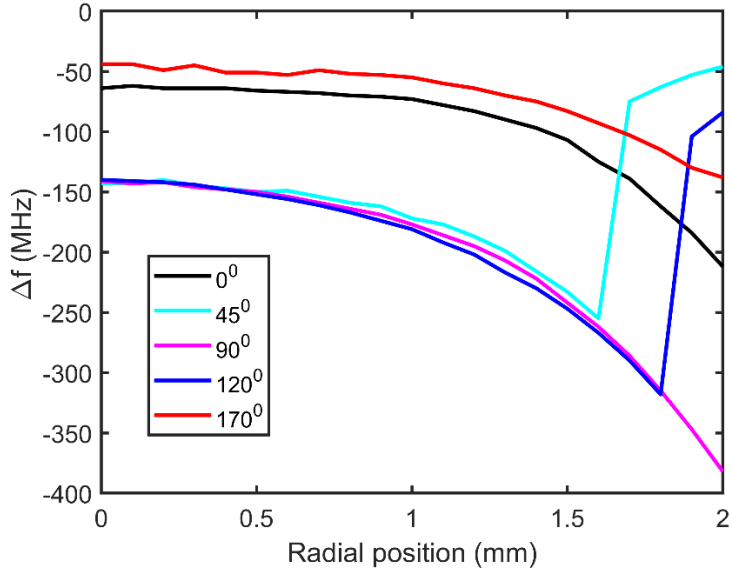

FIG. 10: Change in the frequency of the cavity as a magnet slide towards the edge of the stub with different orientations. The sketches adjacent to the graphs indicate the orientation of the magnet with respect to the stub.

Table 2 summarizes the change in the cavity frequency at different orientations of the magnets at the center and edge of the stub. There is an opposite trend of a frequency shift on those locations of the stubs with orientation less than or equal to 120 degrees and greater than 120 degrees.

Table 2: Summary of the frequency change as the magnet, in contact with the stub, slides towards the edge of the stub with different angles.

| Orientation (°) | $\Delta f\|_{x=0\ mm}$ (MHz) | $\Delta f\|_{x=2\ mm}$ (MHz) | Total f change (MHz) | Total f change (%) |
|---|---|---|---|---|
| 0 | -62 | -212 | -150 | 241 (↓) |
| 45 | -143 | -46 | +96 | 67 (↑) |
| 90 | -141 | -382 | -241 | 171 (↓) |
| 120 | -140 | -84 | +56 | 40 (↑) |
| 170 | -44 | -138 | -94 | 214 (↓) |

IV. Rotation

Another phenomenon to consider during the levitation experiment is the rotation of the magnet. The rotation of the magnet also affects the frequency of the cavity. Fig. 11 illustrates the effect of magnetic rotating at $x = 1.5\ mm$ on the frequency of the cavity. In these simulations, the center of mass of the magnet is fixed at $x = 1.5\ mm$, and the distance between the magnet and the stub is adjusted so that the magnet is always in contact with the stub. In addition, the rotation of the magnet is done manually.

The rotation of the magnet results in an asymmetry double-well like frequency shift profile (see Fig. 11). This asymmetry arises because the part of the magnet that is in contact with the stub moves inwards while making the acute angle and it moves outward while making the obtuse angle to maintain the center of mass of the magnet at $x=1.5\ mm$ (see inset of Fig. 11). The center of the well is at 90 degrees. For the acute angle highest shift of ~$200\ MHz$ is observed for angles 50, 60, and 70 degrees. In the case of obtuse angle, the maximum of $220\ MHz$ of frequency downshift is observed at 100 degrees and is remains reasonably constant for angles 100-140 degrees. For the angle greater than 140, an increasing trend of the frequency shift is observed.

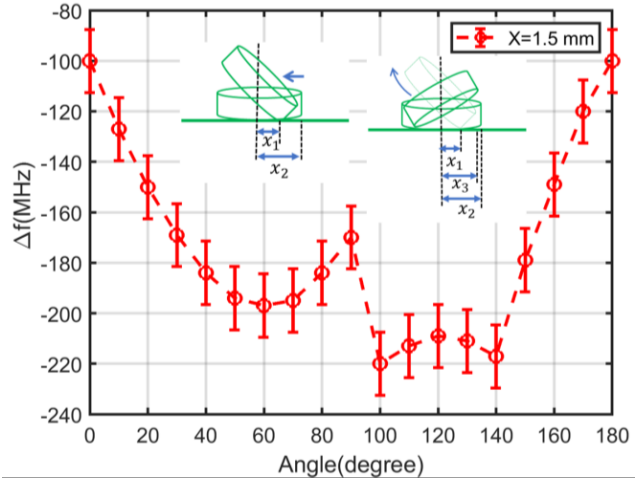

FIG. 11: Effect of magnetic rotation at the edge of the stub on the frequency of the cavity. The average error bar of 25 MHz is implemented in the graph.

V. Conclusion

Figure 12 generalized shift pattern observed in the FEM simulations. The blue spectra represent the bare cavity frequency. A magnet placed on the stub pulls the frequency of the cavity (orange spectra) down at least by 50 MHz. A magnet, when lifted from the surface of the stub, always increases the cavity's frequency irrespective of its position and angle on the stub (represented by the purple arrow). The effect is drastic at the edge of the stub. The frequency upshift due to magnetic levitation is in the hundred orders larger in magnitude than the frequency upshift when the magnet makes a small angle with the stub (light blue arrow). The frequency goes down when the magnet is pushed off towards the edge of the stub (green arrow).

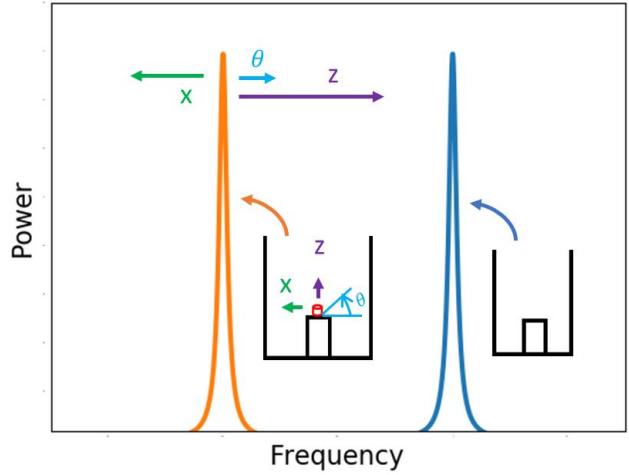

FIG. 12: Generalized frequency shift pattern for key phenomenon during the levitation experiment.

B. Lump Element Model

Now we know the amount of frequency shift that can be estimated from the FEM simulation. This section uses an analytical model that establishes the relationship frequency of the cavity with the strength of the magnet.

A magnet resting on the surface of the stub changes the resonance frequency of the bare cavity. The magnet's perturbation to the field of the cavity can be modeled by using a lumped circuit model [34]. In this model, the frequency of the cavity is given by:

$$f = \frac{1}{2\pi\sqrt{L_0 C}} \tag{3}$$

where $L_0 = \frac{\mu h_s}{2\pi} \ln\left(\frac{r_c}{r_s}\right) = 0.14$ µH and $C$ are the inductance and capacitance of the of the cavity, respectively. Here, $r_c$ denotes the radius of the cavity, $r_s$ the radius of the stub, and $h_s$ the height of the stub. For a cavity with a long cylindrical section between the stub and the open end of the cavity, the electric field is concentrated in the coaxial region, and the exact expression for the capacitance is not known [41]. We assume that a magnet placed near the top of the stub ($z <$ 1 mm), where the electric field is strong, changes the effective capacitance of the cavity according to $C = C_0 + C_1$, where $C_0$ quantifies the capacitance of the bare cavity and $C_1$ represents the capacitive contribution of the magnet. The $C_1$ can be modelled as the interaction of the magnet with the energy density of the cavity falling approximately by $e^{-2\beta z}$ into the waveguide section of the cavity, where $\beta = \sqrt{(\frac{2\pi}{\lambda})^2 - (\frac{2.41}{r_c})^2}$, $\lambda$ is the resonance wavelength [37]. A Taylor expansion of (3) about the bare-cavity resonance frequency, $f_0 = 10.04$ GHz, gives $f^2 = f_0^2(1 - C_1/C_0)$. Using the approximations for capacitance given above we can express the frequency shift as:

$$\frac{f_0^2 - f^2}{f_0^2} \propto e^{-2\beta z} \tag{4a}$$

$$z \propto -\frac{\ln\left(\frac{f_0^2 - f^2}{f_0^2}\right)}{2\beta} \tag{4b}$$

In equation (4), we see how the levitation height and frequency shift are related to one another. We can also relate the levitation height of the magnet to the balance of forces in equilibrium. The levitation force depends quadratically on the strength of the magnet as $F_{Lev} \propto m^2/z^4$, where $m = \frac{B_r V}{\mu_0}$ is the magnet's moment, $z$ is the levitation height as measured between the magnet's center of mass and the superconducting surface, $B_r$ is the remanence of the magnet, and $V$ is the volume of the magnet [42]. For magnetic levitation, the levitation force must counterbalance the gravitational force, $F_G$. The value of $F_G = \rho V g$ is constant for the magnets considered here. Upon steady-state levitation we obtain [43]:

$$z \propto B_r^{\frac{1}{2}} \tag{5}$$

Combining the relationships in (4) and (5) we obtain:

$$-\frac{\ln\left(\frac{f_0^2 - f^2}{f_0^2}\right)}{2\beta} \propto B_r^{\frac{1}{2}} \tag{6}$$

Equation (6) established a relationship between the change in the resonance frequency of the cavity due to the position of the magnet with the strength of the magnet.

C. Quality Factor

The figure of merit most commonly ascribed to cavity resonators is the quality factor (Q), defined as:

$$Q = \frac{f_r}{\Delta f} \tag{7}$$

where $f_r$ is the frequency of the resonant mode of the cavity and $\Delta f$ is the full width at half maximum (FWHM) bandwidth. This expression for the quality factor is in reference to the loaded $Q$ ($Q_L$), which takes losses at the input-output couplers into account. The intrinsic $Q$, $Q_0$, is the quantity that describes the maximum potential for quality factor given an ideal coupling strength [33] and is defined as:

$$Q_0 = Q_L(1 + \beta) \tag{8}$$

$\beta$ is a unitless parameter that describes the strength of the coupling between the input-output signal and the fields inside of the cavity. Note that when doing measurements in reflection ($S_{11}$) only one $\beta$ is needed. However, because a transmission measurement ($S_{21}$) requires two input-output couplers, a second $\beta$ must be calculated at the second coupler in $S_{22}$. This alters Eq. (4) as the following:

$$Q_0 = Q_L(1 + \beta_1 + \beta_2) \tag{9}$$

The actual method for calculating $\beta$ will be discussed in a experiment section.
If we look at $Q$ as a function of energy, the expression for $Q_0$ can be written as:

$$Q_0 = \frac{E_{Stored}}{P_{dissipation\ per\ cycle}} = \frac{2\pi f_r \mu_0 \int_V |H|^2 dV}{R_s \int_S |H|^2 dS} \quad (10)$$

Where $\mu_0$ is the vacuum permeability, $H$ is the magnetic field, and $R_s$ is the surface resistance. When experiments are done in a dilution refrigerator at a base temperature below Tc, the surface resistance will drop into the nΩ range and increase $Q_0$ significantly. This means that with an appropriate coupling coefficient, $\beta$, the loaded Q will also increase.

## 5. Stable Magnetic Levitation: Challenges

An example configuration of the 3D microwave cavity that has been used in the discussion of this paper is shown in Fig. 13. This cavity is made from 6061 aluminum, a type I superconductor. A magnet is put on the stub of the cavity. The aspect ratio of the stub and magnet is 4:1 (ratio of the radius of the superconductor to the radius of the magnet). The dimension of the wall and floor of the cavity are large compared to the dimension of the magnet. Due to its large size, the wall and the cavity floor can be considered a semi-infinite plane.

A hybrid technique for the potential energy calculation is developed here by combining the two-loop [44] and the mirror method [45]. Two-loop model is used for the interaction between the magnet and stub. Whereas in the case of the magnet and wall interaction, the mirror method is used. The potential energy for such a cavity-magnet system will be:

$$U_{total} = U_{two-loop} + U_{Mirror} + U_{Gravitaitonal} \quad (11)$$

$$U_{total} = \frac{\mu_0 I m}{2\pi[(R_S + r)^2 + h^2]^{\frac{1}{2}}} \left[ \frac{R_S^2 - r^2 - h^2}{(R_S - r)^2 + h^2} \cdot E(k) + K(k) \right] + \frac{\mu_0 m^2}{4\pi} \frac{(1 + sin^2\theta)}{16(d - x)^3} + Mgh \quad (12)$$

Here, $2h$ is the distance between the magnet and its diamagnetic image, $I$ is the response supercurrent.

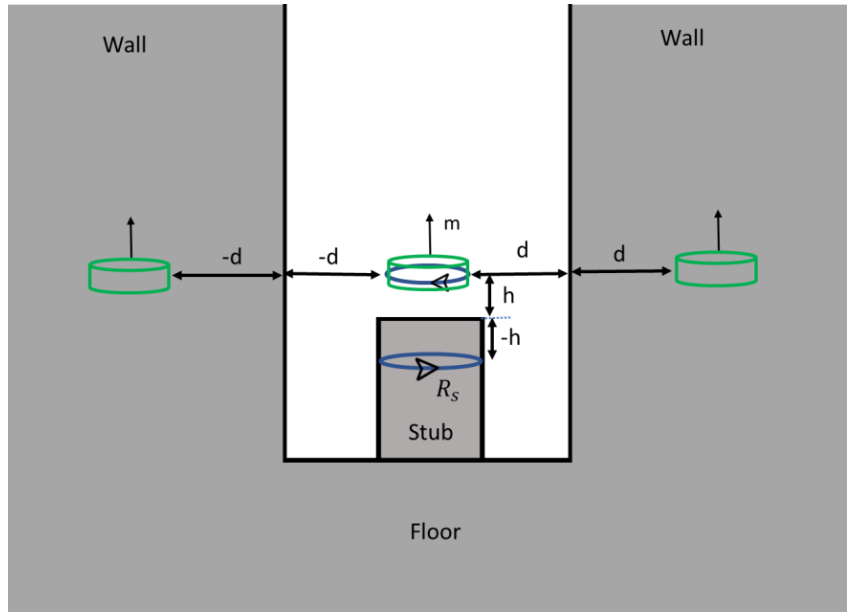

FIG. 13: A 3D cavity configuration. This configuration has the dimension same as the cavity we are using in our experiment.

Figure 14 shows a contour plot of the potential energy of the 3D cavity. The origin of the coordinate is considered at the center of the stub. The calculation is done in the positive x-direction.

In the calculation, the minimum potential energy of the cavity is found, as shown in Fig. 14, at $(x, z = 3.5\ mm, 2.1\ mm)$. In other words, the minimum potential energy is at the center of the cavity gap. One advantage of knowing this local potential energy minima is that we see a magnet on the surface of the stub will be pushed towards this point when the stub and the cavity go into the Meissner transition. Therefore, it will fall on the bottom of the cavity. There needs to be some external resistance to keep the magnet on the stub [46]. One can fit a plastic sleeve on the stub for that purpose. The drop in quality factor would be the price to pay in exchange for that.

There will be two ways to pursue stable magnetic levitation on a 3D cavity configuration stub. First, one can change the gap of the cavity. The second option will be a modification in the configuration and strength of the magnet. The following sub-section discusses these two ways of achieving stable magnetic levitation.

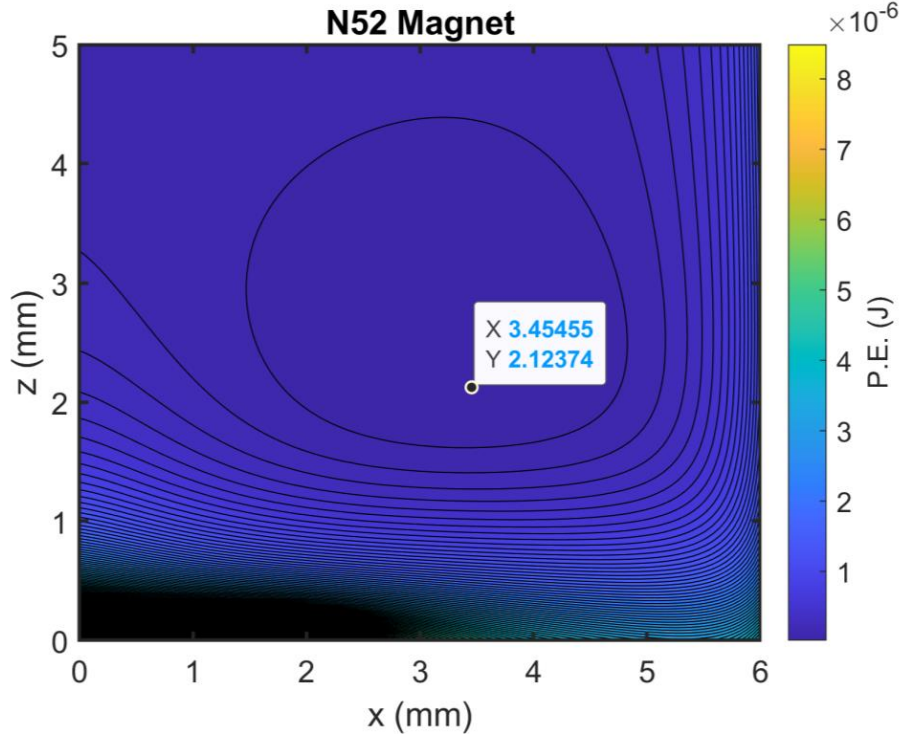

FIG. 14: Potential energy calculated as a function of radial and vertical position of the magnet. The calculation is done for the positive direction from the center of the cavity.

I.  Cavity Configuration

The first option for the stable magnetic levitation within a 3D superconducting cavity is to reduce the gap between the stub and the wall of the cavity. The force from the wall to the magnet increase with $\propto \frac{1}{z^4}$ [47]. Fig. 15 calculates minimum potential in the radial distance for the gap size from 5-2 mm using Eq. (12). When the gap is 3 mm, the magnet on the stub sees the least potential in the center of the stub and the wall. In quest of going into the least energy point, it falls on the gap of the cavity. If the gap is reduced to $\leq 2.5$ mm, the minimum potential point lies well within the stub. At this gap, the magnet can stably levitate on the stub without falling into the gap.

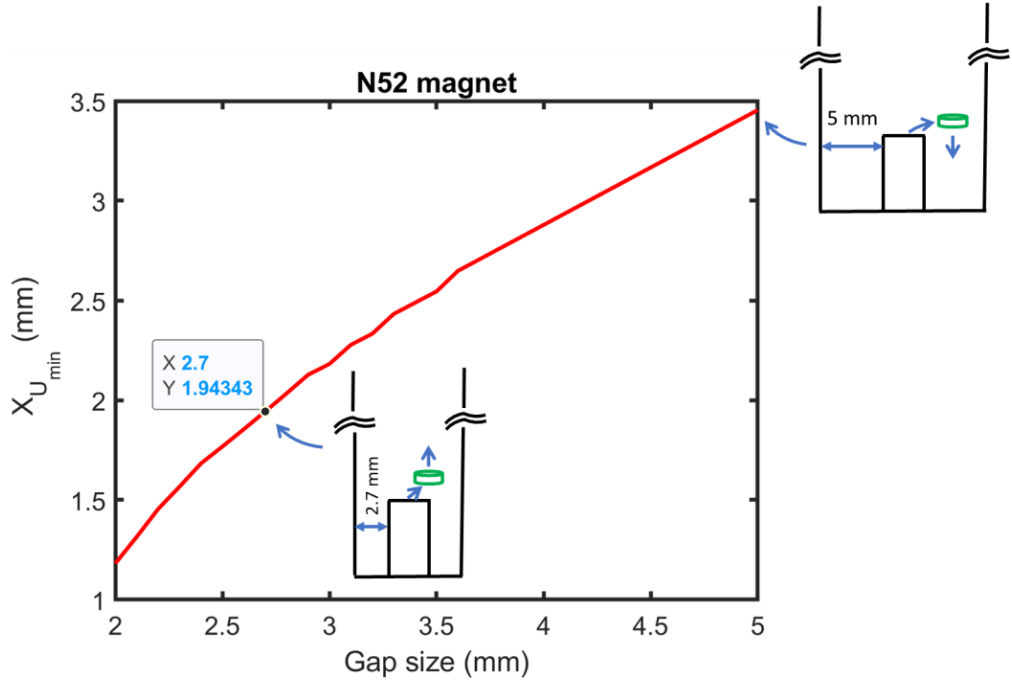

FIG. 15: Variation in the gap size of the 3D cavity configuration for the stable magnetic levitation.

II.  Configuration of the Magnet

Equation (12) shows three aspects of the magnet that plays role in the total potential energy of the cavity-magnet interaction. This subsection of the paper discusses the effect of a magnet's strength, orientation, and size in stable magnetic levitation.

i.  Strength

The strength of the magnet refers to its remanence field. A study of the remanence field effect on the radial and vertical position of magnetic levitation is presented in Fig. 16. The magnet is axially and radially magnetized. The remanence field is varied from 0 to 2 T.

For both axially and radially polarized magnets, the minimum energy lies in the gap of the cavity. In such a cavity configuration, the magnet then falls on the gap and hence on the bottom of the cavity. Although the magnet will be pushed towards the gap, the radially polarized magnets have improved the location of the levitation. Such magnets moved the radial position from 3.5 mm to 2.5 mm. Remember, the radius of the stub is 2 mm. This point is only 0.5 mm outside of the stub.

In the case of the vertical levitation distance, an increasing trend with the strength of the magnet is seen for both cases of magnetization. The radially polarized magnets have shown it at the lower height. The reason is the induction of a weaker diamagnetic image when the magnetic moment is parallel to the superconducting axis providing a weaker lifting force.

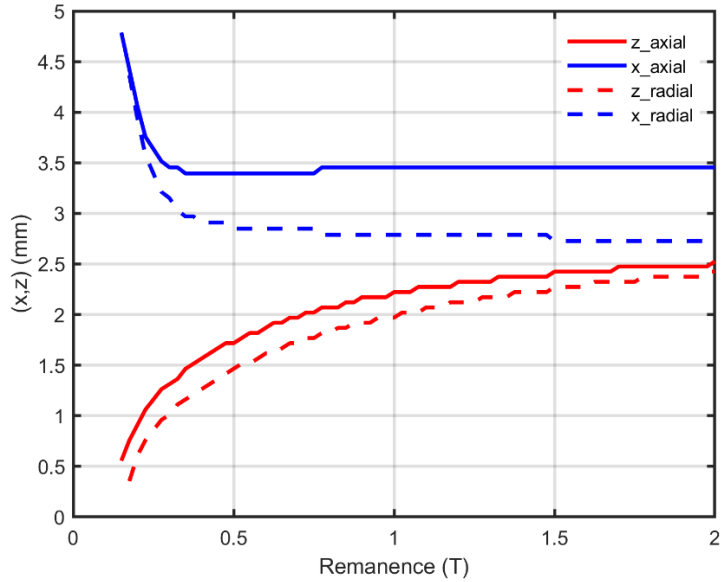

FIG. 16: Study of radial and vertical position of magnetic levitation for an axially and radially magnetized magnet with strength ranging from 0 to 2 T.

  ii.  Orientation

The orientation of the magnet is changed from zero degrees to 90 degrees with respect to the stub by keeping the strength of the magnet constant to 1.47 T. The potential energy above the stub is calculated for those angles. The axially polarized magnet is used for the calculations. The vertical and radial position is extracted from the minimum energy state and plotted in Fig. 17.

As the angle of the magnet is changed from zero degrees to 90 degrees, the radial position of the magnet is shifted from 2.8 mm to 3.5 mm. This is a 25 percent push for a magnet in the radial n from its original position. The angles have less effect on the lifting height. During the flip of the magnet (zero degrees to 90 degrees), the levitation height is changed only by 9 percent.

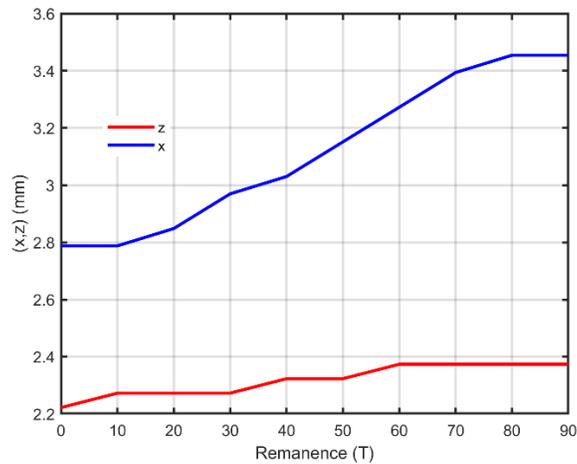

FIG. 17: The location of the magnet moves when the magnet changes its angle during magnetic levitation. The magnet is N52 and is polarized in the axial direction.

iii. Size

The other important factor that plays a role in stable magnetic levitation is the size of the magnet. The variation in the radius or thickness of the magnet changes its volume. The volume itself changes with the square of the radius of the magnet and is proportional to its thickness. If the dimension of the magnet is changed by keeping its strength the same, there will be a change in two important factors that determines the levitation force. First, when the volume of the magnet changes, it changes the field on the surface of the magnet too. Second, the volume of the diamagnetic image varies proportionally to the magnet's volume.

Figure 18 shows the radial position of magnetic levitation as a function of the dimension of the magnet. The radius of the magnet is varied from 0.5 mm to 1.5 mm. Similarly, the thickness is varied from 0.5 mm to 0.75 mm. The change in the radial position is seen for the magnet with a larger thickness. For example, suppose the size of the present magnet that has a radius and thickness of 0.5 mm is scaled up to a radius of 1.5 mm and a thickness of 0.75 mm. In that case, the radial position of magnetic levitation will shift towards the stub (2.5 mm) from the center of the gap (3.5 mm).

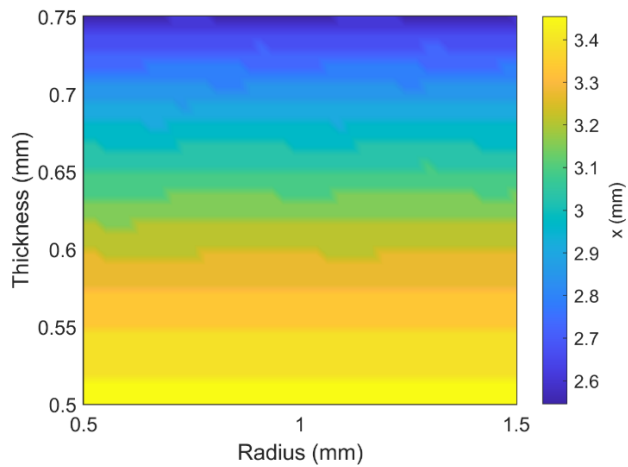

FIG. 18: The present magnet's dimensions (0.5 mm of radius and thickness) are scaled up to see its effect on the radial location of magnetic levitation. The strength of the magnet is kept constant at 1.47 T.

### 6. Experiments: Possibilities

A. Room Temperature measurements

To validate the FEM simulation results and understand the frequency shifts when levitation occurs, the resonance frequency of the cavity is measured as a function of the position of a magnet in the cavity at room temperature. Fig. 19 shows a schematic of room temperature measurements. Measurements are taken by putting a magnet inside a capillary tube and by sealing its end with tape and factoring out the effect of the capillary. The capillary tube with the magnet is positioned at different coordinates inside the cavity by the translational stage of the micrometer. The magnet was held in a dielectric capillary and its position was controlled using micrometer stages.

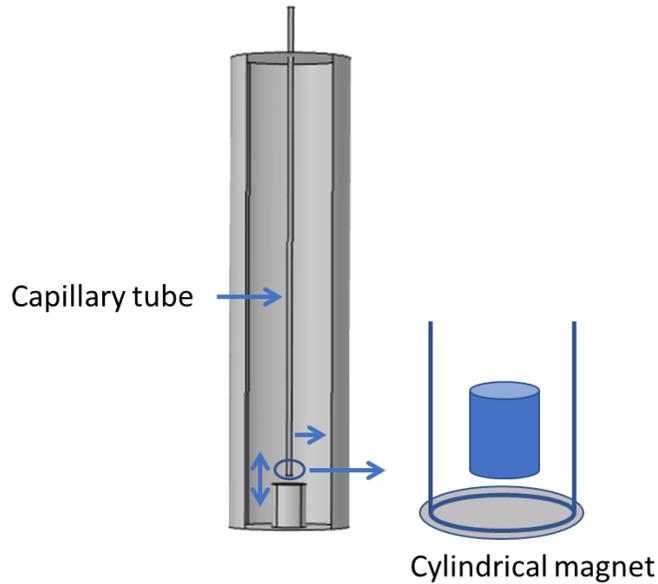

FIG. 19: Schematic of the room temperature measurements. The measurements are performed on and around the stub and bottom of the cavity. A cylinder magnet with a radius of 0.375 mm and a height of 1 mm is used.

First, the stub is scanned by a moving magnet in the radial direction. The idea is to find the symmetry of the curve and hence the origin of the measurements. The change in cavity resonance as a function of the magnet's lateral position is illustrated in Fig. 20. Each curve represents a different vertical position, and the horizontal axis is the radial distance from the cylinder's axis. The cavity resonance is more sensitive to the magnet's position when it is located above the stub ($|x|<2.5$ mm) than above the gap ($|x|>2.5$ mm) with the region of highest sensitivity near the stub's edge ($|x|\sim1.5$ mm). We attribute the asymmetry in the frequency shift with x to machining imperfections.

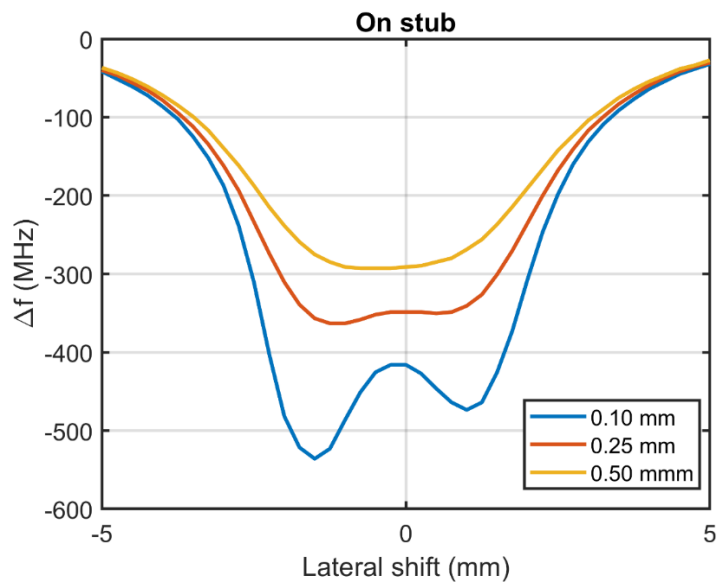

FIG. 20: A cylindrical magnet with a height of 1 mm and a radius of 0.375 mm is placed at different positions above the stub to measure the sensitivity of the cavity. The magnet was held in a dielectric capillary, and its position was controlled using micrometer stages.

Now, the vertical translation of the magnet is done outside of the stub in Fig. 21. This will help make a map of the change in frequency outside of the stub. The magnet is moved from 4 mm above the stub to the bottom of the cavity.

The magnet on and above the stub has shown a similar frequency shift trend that is seen in the above figure. The trend of frequency shift reversed as the magnet got past the stub towards the bottom of the cavity. The magnet plays the opposite role sitting on the stub and bottom of the cavity. The magnet on the stub increases the effective height of the stub. The frequency of the cavity, hence, downshifts. However, on the bottom of the cavity, it tends to raise the floor of the cavity. Thereby reducing the effective height of the stub. This, in contrast, results in the upshift of the frequency.

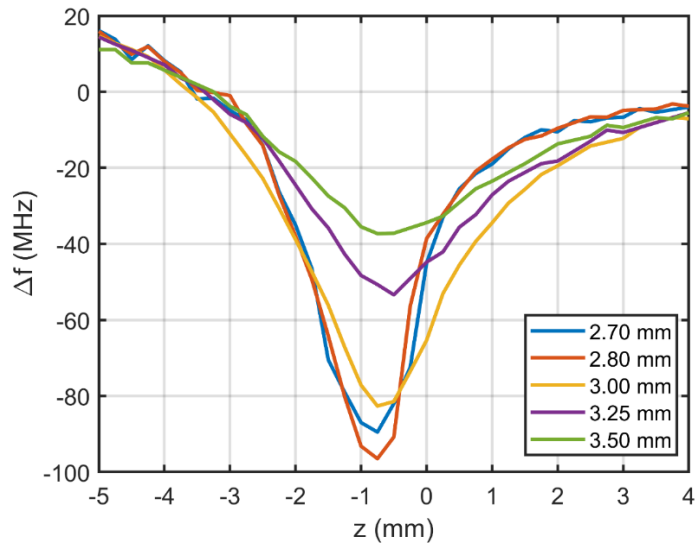

FIG. 21: The magnet is vertically translated at different locations outside of the stub. The starting position of magnetic translation is 4 mm above the stub, and it ends up on the bottom of the cavity (-5 mm).

Figure 22 illustrates the change in the resonance frequency of the cavity in the region between the stub and wall of the cavity. The magnet is laterally moved from the wall towards the stub. According to the trend of the frequency shift, the region can be divided into three parts: (a) $0 \leq z < 2.5$ mm, (b) $z \cong 2.5$ mm, and (c) $z > 2.5$ mm.

The first region (a) is the region that lies close to the bottom of the cavity. In this region, the magnet interacts mainly with the rf magnetic field of the cavity. Such interactions result in a frequency upshift from the empty cavity frequency ($\Delta f=0$). A maximum of 25 MHz frequency upshift is observed when the magnet is on the bottom and close to the stub of the cavity. The region (b) is halfway between the stub and bottom of the cavity. The sensitivity is least in this region among all three regions. The availability of the rf fields is low. In this part of the cavity, one can even hide the magnet without seeing significant changes in the frequency of the cavity. The push of the magnet from the wall towards the stub, in the region (c), turns around the change frequency shift that is seen at the bottom of the cavity. The magnet in this region interacts with the cavity's electric field.

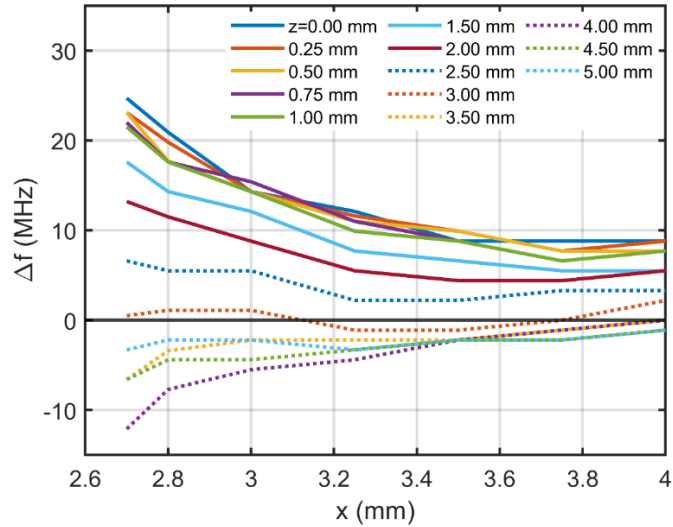

FIG. 22: The region between the stub and wall of the cavity is probed by the magnet. Here, Δf=0 is the frequency of the bare cavity.

The above room temperature measurements strengthen the frequency shift patterns for the magnetic levitation from the stub and bottom of the cavity. Moreover, it also strengthens the FEM simulation calculation done in section 4 (A).

B. Cryogenic Measurements

Magnetic levitation above a type-I superconductor is due to the Meissner effect. For that effect to see a superconducting has to cool down below its critical temperature. For aluminum, such temperature is 1.2 K. To achieve such a low temperature needs a dilution refrigerator. In this section of the paper, the main discussion will be on different aspects of the cryogenic measurements.

I. Coupling Mechanisms

Different coupler geometries (see Fig. 23) may be used to couple a signal to the magnetic field (loop coupler) or the electric field (straight pin coupler). Coupler placement has a large effect on coupling as well. In the "straight probe", the coaxial cable is transmitting a mode with a radial electric field that extends from the center conductor to the grounded outer cylinder. When the end of the transmission line is left open-circuit as shown in Fig. 23 (a), the electric field lines direct themselves towards a ground plane somewhere on the interior walls or stub of the cavity. A small overlap of this applied electric field with the cavity mode is sufficient to excite the low-loss cavity. For the 10 GHz mode, the electric field is concentrated at the tip of the stub and decays exponentially as it moves towards the open top of the cavity. The magnetic field is concentrated at the cavity floor around the stub. Fig. 23 (b) illustrates coupling to the magnetic field where the central conductor is bent around and shorted to the shielding. Current owing back and forth through this semi-circular section produces a magnetic field as shown which, if oriented properly, overlaps sufficiently with that of the cavity mode, thus exciting the cavity.

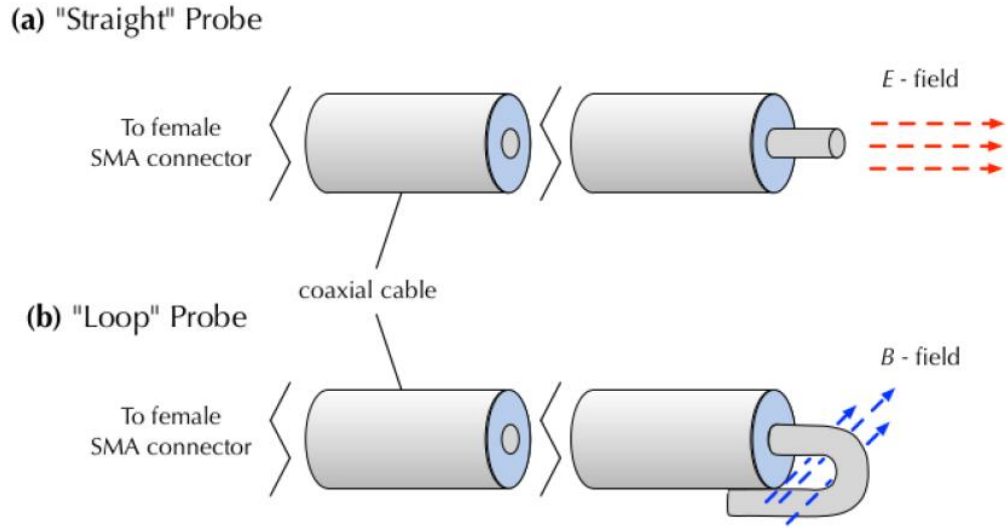

FIG. 23: Pin and loop configurations that couple to $E$ and $B$, respectively. The pin configuration consists of a piece of bare wire soldered onto the end of a SubMiniature version A (SMA) female connector and couples to the electric field like that of an electric dipole antenna. In the loop configuration, a bare wire is soldered onto the end of the same type of SMA connector used for the pin configuration but is looped back around such that it makes contact with the cavity wall and forms a closed circuit [3].

II. Testing techniques

For the experimental measurements, an HP 8720C Vector Network Analyzer (VNA) is used to find the resonance and an approximate value for the quality factor. The VNA is a device that measures the amplitude and phase response of a device under test as a function of the applied frequency. Fig. 24 shows the measurement configuration for transmission and reflection measurements.

There would be the case where the resolution VNA's is not high enough. In that situation, one other measurement technique is to use the HP 8593B electrical spectrum analyzer (ESA) in combination with a signal generator (see Fig. 24). By using the peak hold functionality of the ESA, we can sweep the signal generator around the resonance to get a much higher resolution view of the cavity's response.

Despite being able to get a higher resolution by utilizing the ESA, cavities at cryogenic temperatures that have transitioned to a superconducting state may have quality factors much higher. To measure the quality factor of these cavities, we use a technique called the ring-down measurement, shown in Fig. 25.

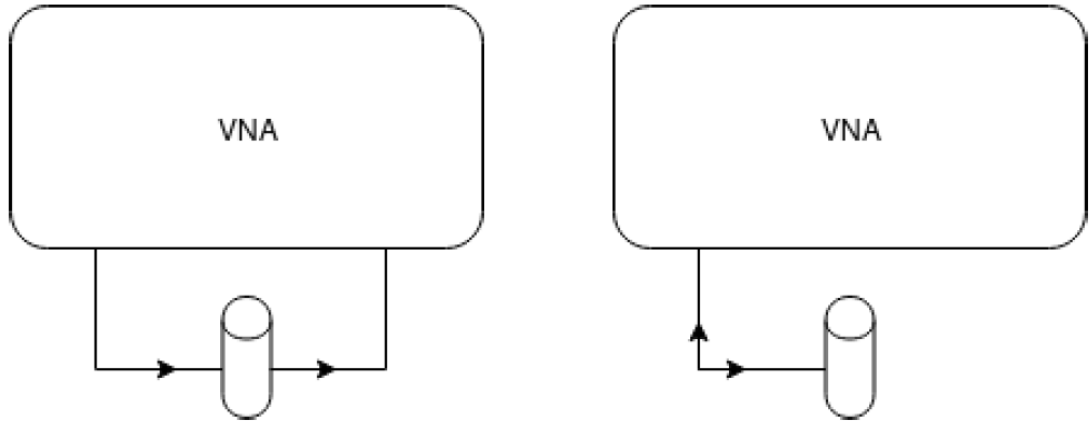

FIG. 24: (Left) Transmission (S21) measurement. A probe signal is swept over a range of frequencies with the response amplitude being greatest on resonance. (Right) Reflection (S11) measurement. The probe signal sweeps over a frequency range and the on-resonance response is a dip because of the energy being absorbed.

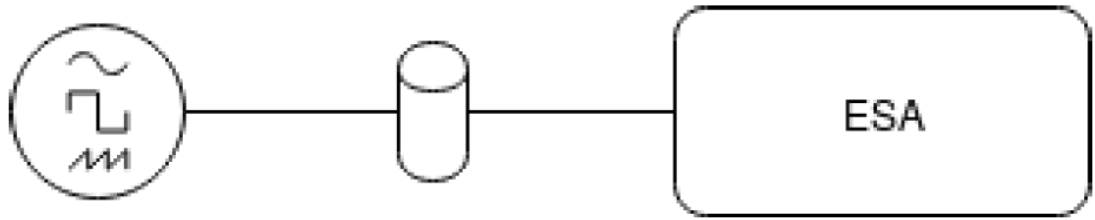

FIG. 25: Turning the ESA into a high-resolution network analyzer. The ESA is set to "peak hold" and the frequency of the probe signal is slowly moved around the resonance of the cavity to increase the resolution of the resonance shape. This measurement may also be done in reflection by placing a three-port circulator at the input of the cavity.

The ring-down measurement uses a signal generator to probe the cavity and an oscilloscope that is triggered by a separate function generator. The cavity is probed near the room temperature resonance and is modulated in frequency space until a response is seen on the oscilloscope. When the cryogenic resonance (which is likely to differ slightly from the room temperature resonance due to thermal contraction of the cavity) is found, the frequency modulation is turned off and the probe signal is pulsed. The resulting data is shown in Fig. 26. By fitting the decay trace to an exponential, $f(t) = e^{-\frac{t}{\tau_L}}$, the decay time can be recovered and related to the cavity's intrinsic quality factor via:

$$Q_0 = 2\pi f_0 \tau_d (1 + \beta) \qquad (29)$$

Where $f_0$ is the resonance frequency and $\beta$ is the coupling coefficient and is found from the ratio of the peak amplitudes from the pulsed trace by:

$$Q_0 = \frac{1}{2\sqrt{\frac{P_f}{P_e} - 1}} \tag{30}$$

Where $P_f$ and $P_e$ are the powers that do and do not couple into the cavity, respectively. It is important to note that $\beta$ can only be found by doing these measurements in reflection. It is, however, possible to get a value for the loaded $Q$, $Q_L$, in transmission [33, 34].

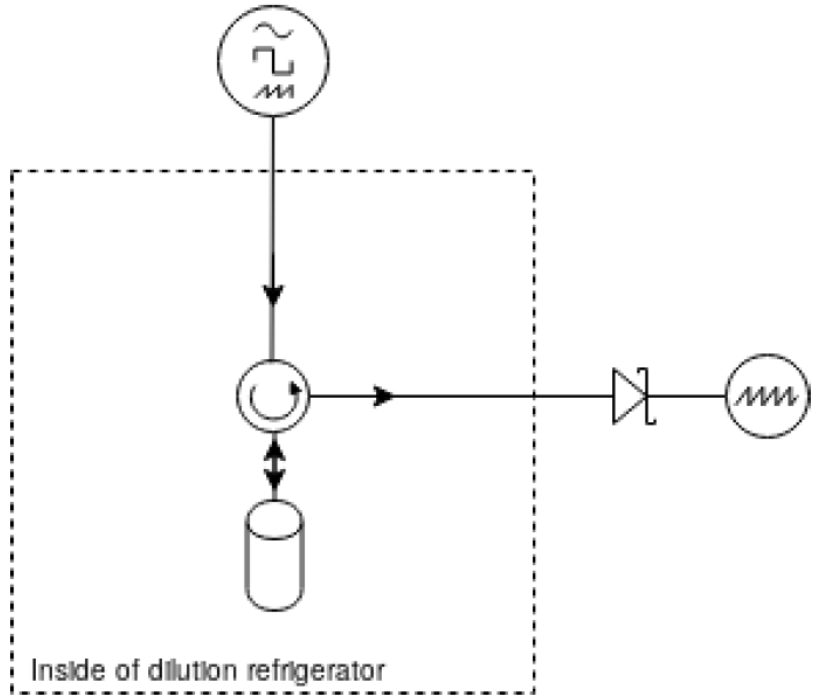

Fig. 26: Schematic of the ring-down measurement. A signal generator is fed into a cryogenic three-port circulator, and the reflected signal is sent out of the dilution refrigerator and through a Schottky diode to convert the power to a voltage. The response is then looked at on an oscilloscope.

Cryogenic experiments are done in a cryogen-free Oxford DR200 dilution refrigerator. The dilution refrigerators consist of a series of plates, all of which serve to cool the experimental chamber down into the tens of milli-Kelvin region which is well below the required transition temperatures of both Niobium (9 K) and Aluminum (1.2 K). The experimental chamber is cylindrical with an inner diameter of 15 cm and a usable inner height of 20 cm, allowing for multiple cavity experiments per cycle. Microwave access to the experimental chamber is granted by custom-made SubMiniature type A (SMA) feed-throughs with access ports on a vacuum-tight angle at the top of the refrigerator that extends down to the base plate. The process to cool the systems into sub-100 mK temperatures is as follows:
1. Place experiment(s) on the base plate and attach microwave input/output lines
2. Replace the four cans that attach to the various plates on the system and begin pumping the Outer Vacuum Chamber (OVC)
3. Let turbo pump get the vacuum in OVC down to $10^{-4}$ mbar
4. Attach helium leak checker to OVC pump line and spray balloon-quality helium around seams of the system to verify that there are no significant leaks

5. Begin pre-cool cycle. This utilizes a mixture of He3 and He4 gas and can cool the base plate to around 4 K over 24-48 hours

6. The full cooldown cycle finishes the cooling by utilizing a pulse tube refrigerator (PTR) and several heat exchangers to drop the base plate down into the sub-100 mK region. This process can take an additional 12-24 hours

As this is a closed system, the expensive mixture of helium gas used in the cooling cycles should all remain in the system during the process. Small leaks are possible, and the mixture is kept pure by pumping the gas through a cold trap in a liquid nitrogen bath. The cold trap is filled with charcoal, and any leaks from the environment surrounding the gas lines will be trapped as anything in the line beyond helium will liquify and be absorbed by the charcoal at liquid nitrogen temperatures (≈70 K).

### III. Analysis of the Microwave Spectra

A microwave spectra (transmitted or reflected) would be a great source for the characterization of magnetic levitation within the microwave cavity when there is no visual access inside the cavity. Two important informations will help identify possible phenomena during magnetic levitation: resonance frequency and quality factor of the cavity. The resonance frequency is obtained from the spectra by extracting the maximum (or minimum) frequency from the transmitted (or reflected) signal. Frequency is directly proportional to the energy. Hence, the resonance frequency is related to the maximum energy a cavity can store. Another important piece of information is the full-width half maximum (FWHM). It is obtained by subtracting (or adding) 3 dBm from the maximum amplitude. The width of the spectra at that amplitude is called the FWHM. It is related directly to the losses of the cavity. The quality factor is the ratio of resonance frequency to the FWHM.

Fig. 27 is an example spectrum that is collected at 5 K. The resonance frequency of this particular spectra is 10.039 GHz and FWHM of 5 MHz. Hence, the quality factor of the cavity at 5 K will be 1985.

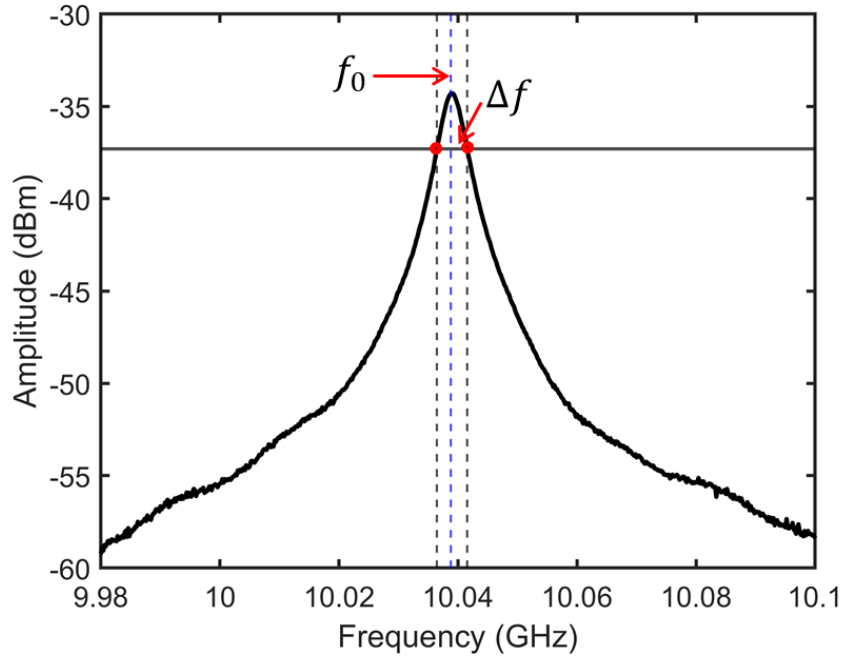

FIG. 27: A microwave spectra of a cavity-magnet system collected at 5 K.

## 8. Summary

The microwave cavity has been utilized as a versatile research platform. The high-Q cavity is used for dark matter detection. It is coupled with an oscillator and is used for sensing, gravitational wave detection, and quantum mechanical experiments. Recently, a novel mechanical perturbation of permanent magnets within the microwave cavity has been realized [42,48]. Such a cavity-magnet system could be an alternative tool for the qubit readout and dark matter detection. Moreover, the powerless transduction between the gravitational and electromagnetic waves was proposed by Gertsenshtein [49]. Here the magnetic field stored between the levitated system plays a part in time-reversal transduction between the vector and the tensor field [50]. The high-Q cavity with a high magnetic gradient of the levitating magnet could be an alternative tool to realize this effect.

In this paper, magnetic levitation within the microwave cavity is characterized using the FEM simulations, Lump-element model, and analysis of the total quality factor of the cavity. A hybrid model is developed by combining the two-loop model and mirror method to calculate the total potential energy for the levitating magnet. A possible scenario for stable levitation by changing the gap of the cavity and shape of the magnet is discussed in brief. The room temperature measurements and procedure and technique for the cryogenic measurements are also presented.